\documentclass[twocolumn,showpacs,preprintnumbers,prd,superscriptaddress,nofootinbib]{revtex4}
\usepackage{doi}%<----------
\usepackage{hyperref}
\hypersetup{colorlinks,linkcolor={blue},citecolor={red},urlcolor={cyan}} 
\hypersetup{
  colorlinks=true,        % false: boxed links; true: colored links
  linkcolor=blue,         % color of internal links
  citecolor=cyan,         % color of links to bibliography
}

\usepackage{bm}
\usepackage{latexsym}
\usepackage{dcolumn}
\usepackage{amsmath,amsfonts,amssymb}
\usepackage{physics}
\usepackage{amsthm}
\usepackage{color}
\usepackage{xcolor}
\usepackage{graphicx,float}
\usepackage{comment}
\usepackage{hyperref}
\usepackage[normalem]{ulem}
\usepackage{enumitem} 
\usepackage{ulem}

\usepackage{pdfpages}
\usepackage{epstopdf}
\usepackage[utf8]{inputenc}
\def\be {\begin{equation}}
\def\ee {\end{equation}}
\def\bea {\begin{eqnarray}}
\def\eea {\end{eqnarray}}
\def\bc {\begin{center}}
\def\ec {\end{center}}
\def\bfg {\begin{figure}}
\def\efg {\end{figure}}
\def\bi {\begin{itemize}}
\def\ei {\end{itemize}}

\def\beq{\begin{equation}}
\def\eeq{\end{equation}}
\def\br{\begin{eqnarray}}
\def\er{\end{eqnarray}}
\newcommand{\eel}[1] {\label{#1}\end{equation}}

\newcommand{\bdm}{\begin{displaymath}}
\newcommand{\edm}{\end{displaymath}}

\usepackage{orcidlink}
\begin{document}

\title{
Spontaneous wave function collapse from non-local gravitational self-energy}

\author{Kimet Jusufi\,\orcidlink{0000-0003-0527-4177}}
\email{kimet.jusufi@unite.edu.mk}
\affiliation{Physics Department, State University of Tetovo, Ilinden Street nn, 1200, Tetovo, North Macedonia}
\author{Douglas Singleton\,\orcidlink{0000-0001-9155-7282}}
\email{dougs@mail.fresnostate.edu}
\affiliation{Physics Department, California State University, Fresno, CA 93740}

\author{Francisco S.N Lobo\,\orcidlink{0000-0001-9155-7282}}%
\email{fslobo@fc.ul.pt}
\affiliation{Departamento de F\'{i}sica, Faculdade de Ci\^{e}ncias da Universidade de Lisboa, Campo Grande, Edif\'{\i}cio C8, P-1749-016 Lisbon, Portugal}
\affiliation{Instituto de Astrof\'{\i}sica e Ci\^{e}ncias do Espaco, Faculdade de Ci\^encias da Universidade de Lisboa, Campo Grande, Edif\'{\i}cio C8, P-1749-016 Lisbon, Portugal}%

\date{\today}

%%%%%%%%%%%%%%%%%%%%%%%%%%%%%%%%%%%%%%%%%%%%%%%%%%%%%%%%%%%%%%%%%%
\begin{abstract}
 We incorporate non-local gravitational self-energy, motivated by string-inspired T-duality, into the Schrödinger–Newton equation. In this framework spacetime has an intrinsic non-locality, rendering the standard linear superposition principle only an approximation valid in the absence of gravitational effects. We then invert the logic by assuming the validity of linear superposition and demonstrate that such superpositions inevitably become unstable once gravity is included. The resulting wave-function collapse arises from a fundamental tension between the equivalence principle and the quantum superposition principle in a semiclassical spacetime background. We further show that wave functions computed in inertial and freely falling frames differ by a gravitationally induced phase shift containing linear and cubic time contributions along with a constant global term. These corrections produce a global phase change and lead to a spontaneous, model-independent collapse time inversely proportional to the mass of the system.
\end{abstract}
%%%%%%%%%%%%%%%%%%%%%%%%%%%%%%%%%%%%%%%%%%%%%%%%%%%%%%%%%%%%%%%%%%

\maketitle

%%%%%%%%%%%%%%%%%%%%%%%%%%%%%%%%%%%%%%%%%%%%%%%%%%%%%%%%%%%%%%%%%%
\section{Introduction}
%%%%%%%%%%%%%%%%%%%%%%%%%%%%%%%%%%%%%%%%%%%%%%%%%%%%%%%%%%%%%%%%%%

Quantum mechanics is the foundational theory governing the behavior of physical systems at microscopic scales. The fundamental equation of quantum mechanics is the Schr\"odinger's equation  \cite{Schrodinger:1926gei}. Unlike classical mechanics, quantum mechanics is governed by unique principles such as superposition, entanglement, and wave function collapse. However, the nature of wave function collapse remains an open question, giving rise to the measurement problem. This problem concerns whether the collapse of the wave function is a fundamental, physical process or merely an emergent phenomenon that arises from interactions with the environment. The resolution of this issue depends on the interpretation of quantum mechanics.

One of the first models to address wave function collapse was proposed by Ghirardi-Rimini-Weber \cite{Ghirardi:1985mt}. Another particularly interesting approach to the measurement problem is gravity-induced wave function collapse, a concept independently proposed by Di\'osi and Penrose \cite{Diosi:1986nu,Diosi:1984wuz,Diosi:1989hlx,Penrose:1996cv,Penrose:1998dg} and studied further in  \cite{Gasbarri:2017alx,Tomaz:2024rzu,Bose:2024nhv,Trillo:2025kio,Gundhi:2025aaa,Diosi:2022kay,Diosi:2019zkh} and explored very recently in \cite{Hossenfelder:2025uqw}. Their models suggest that gravity plays a fundamental role in the collapse mechanism, and explains why macroscopic objects cannot exist in quantum superposition in the same way as microscopic particles. In contrast to standard quantum mechanics, which treats gravity as an external classical field, the Di\'osi-Penrose (DP) model proposes that superpositions of different gravitational field configurations become unstable and collapse spontaneously.  A particular model of Di\'osi-Penrose  was recently tested, and experimental bounds were obtained using an underground test of gravity-related wave function collapse \cite{Donadi:2020kzc}. In the present work we will discuss how the wave function collapse model of the present work is not subject to the constraints of \cite{Donadi:2020kzc}.

Penrose, in particular, has argued that the collapse of the wave function originates from an inherent conflict between the principle of equivalence and the principle of superposition when a gravitational field is present \cite{Penrose:2014nha,Howl:2018qdl}. His argument follows from the fact that wave functions computed in an inertial frame under a gravitational field and those in a free-fall reference frame are related by an additional phase factor that depends on the gravitational acceleration. This phase includes both linear and cubic time-dependent terms, leading to an ambiguity in defining a well-posed spacetime metric when such states are superposed. Consequently, the gravitational field itself cannot remain in a coherent superposition, leading to an intrinsic breakdown of quantum linearity and ultimately causing wave function collapse. 

In Refs.~\cite{Penrose:2014nha,Howl:2018qdl}, the spacetime background is treated semi-classically. 
In contrast, we study a specific gravitational potential emerging from string T-duality \cite{Sathiapalan:1986zb}, where both the potential and the spacetime are free of singularities. 
String T-duality connects string theories in higher-dimensional spacetimes with inverse compactification radii $R$, transforming quantum numbers as 
$R \rightarrow {R^\star}^2/R$ and $n \leftrightarrow w$, where $R^\star = \sqrt{\alpha'}$ is the self-dual radius, $\alpha'$ the Regge slope, $n$ the Kaluza–Klein excitation, and $w$ the winding number. 
The resulting quantum-corrected propagator remains regular, incorporating an ultraviolet cutoff determined by the zero-point length $l_0 = 2L$ \cite{Padmanabhan:1996ap,Smailagic:2003hm,Fontanini:2005ik}. 
T-duality-inspired models have been applied in the context of black holes and cosmology \cite{Nicolini:2022rlz,Nicolini:2019irw,Gaete:2022ukm,Nicolini:2023hub,Jusufi:2024dtr,Jusufi:2025qgd,Jusufi:2022mir,Millano:2023ahb}, offering a mechanism that might resolve the singularities in both neutral and charged black hole solutions.

This gravity-induced collapse framework provides a potential link between quantum mechanics and general relativity, suggesting that gravity may play a role in the collapse of the wave function \cite{Hossenfelder:2025uqw,Wang:2025rdv}. By employing T-duality, one can formulate a non-local gravitational theory with a regularized gravitational potential, where the particle mass is not strictly localized but rather smeared out and described by a quantum-corrected energy density. One of the implications is that in short distances the spacetime itself becomes inherently not defined due to the presence of an additional quantum uncertainty induced by gravitational effects. Using a regular gravitational potential and non-local gravitational self energy, we further argue that gravity-induced wave function collapse emerges naturally from fundamental spacetime fluctuations. In this approach, one can further construct a regular spacetime geometry which, depending on the mass parameter, can describe a particle or a black hole \cite{Jusufi:2025qgd}. 

It is important to emphasize the key differences between the present framework and the well-known Diósi–Penrose (DP) approach. In the DP model, the gravitational self-energy responsible for collapse is introduced phenomenologically as a measure of the instability between superposed mass distributions. In contrast, in our approach the gravitational self-energy arises directly from a non-local gravitational theory motivated by T-duality, and is therefore not an external input but a derived quantity. Furthermore, the resulting Schrödinger–Newton equation is regularized due to the presence of a zero-point length, removing the short-distance divergences present in the standard formulation. Finally, while the DP framework is often implemented via stochastic modifications of quantum mechanics, the present model is entirely deterministic at the level of the dynamical equation, with collapse emerging from the breakdown of spacetime definiteness at short scales.

The paper is organized as follows. In Sec.~\ref{SectionII}, we introduce the non-local contribution to the gravitational self-energy and derive the corresponding modified Schr\"odinger--Newton equation, emphasizing the role of the minimal length scale and its impact on the effective potential. In Sec.~\ref{SectionIII}, we apply this framework to analyze gravity-induced wave function collapse, examining both the phase structure and the emergent collapse times associated with non-local gravitational effects. In Sec.~\ref{SectionIV} we will discuss how the wave function collapse model of the present work is not subject to the constraints of \cite{Donadi:2020kzc}. In Sec.~\ref{Section:Conclusion}, we summarize our findings and discuss their implications for the interface between quantum mechanics and gravity, as well as potential directions for future research. Throughout this work, we set $c=1$, unless otherwise specified.

%%%%%%%%%%%%%%%%%%%%%%%%%%%%%%%%%%%%%%%%%%%%%%%%%%%%%%%%%%%%%%%%%%
\section{Schr\"odinger--Newton equation with gravitational self energy }
\label{SectionII}
%Regular spacetime geometry in T-duality}
%%%%%%%%%%%%%%%%%%%%%%%%%%%%%%%%%%%%%%%%%%%%%%%%%%%%%%%%%%%%%%%%%%

%%%%%%%%%%%%%%%%%%%%%%%%%%%%%%%%%%%%%%%%%%%%%%%%%%%%%%%%%%%%%%%%%%	
\subsection{Schr\"odinger--Newton equation in T-duality} 
%%%%%%%%%%%%%%%%%%%%%%%%%%%%%%%%%%%%%%%%%%%%%%%%%%%%%%%%%%%%%%%%%%	

We shall closely follow the recent work \cite{Jusufi:2025qgd}, in which the non-local gravitational energy was incorporated into the spacetime geometry. 
To proceed, we recall the quantum-corrected static interaction potential derived from the field theory with the path integral duality proposed in \cite{Nicolini:2019irw}. 
In particular, the momentum-space massive propagator induced by the path integral duality is given by \cite{Nicolini:2022rlz,Nicolini:2023hub}
\begin{equation}
G(k)
= -\frac{l_0}{\sqrt{k^2+m_0^2}}\, K_1 \!\left(l_0 \sqrt{k^2+m_0^2}\right)
\end{equation}
with $l_0$ being the zero-point length and $K_1\!\left(x\right)$ is a modified Bessel function of the second kind. Without loss of generality, we consider the massless propagator case, i.e. $m_0=0$. 
Specifically, we have two cases: at small momenta, we obtain the conventional massless propagator $G(k) = -k^{-2}$. At large momenta, the exponential suppression is responsible for curing UV divergences \cite{Nicolini:2022rlz,Nicolini:2023hub}. 
We briefly recall the origin of this potential. In string theory, T-duality relates theories with compactification radius $R$ to those with radius $R^\star = \sqrt{\alpha'}$, where $\alpha'$ is the Regge slope. This duality implies a minimal length scale $l_0=2 \pi \sqrt{\alpha'}$ \cite{Smailagic:2003hm,Fontanini:2005ik}. The path-integral duality approach \cite{Nicolini:2019irw} modifies the propagator to $G(k) = -l_0 K_1(l_0\sqrt{k^2})/\sqrt{k^2}$, which at large $k$ suppresses UV divergences. Fourier transforming the static limit yields the regularized potential \eqref{potential}.
Consider a static external source $J$ which consists of two point-like masses, $m$ and $M$, at relative distance $\vec{r}$, then the potential is given by \cite{Nicolini:2019irw}
\begin{eqnarray}\notag
V_G(r)
&=& -GM\, \int \frac{d^3 k}{{\left(2\pi\right)}^3}\; { G_{F}(k)|}_{k^0=0}\; 
 \exp\!\left(i \vec{k} \vec{r}\right) \\
 &=&
 -\frac{GM}{\sqrt{r^2 + l_0^2}}. \label{potential}
\end{eqnarray}
Using Poisson's equation gives an energy density function for the bare matter as
\cite{Nicolini:2019irw}
\begin{equation}
\rho^{\rm bare}(r)= \frac{1}{4\pi G}\nabla^2 V_G(r)=\frac{3 l_0^2 M}{4 \pi \left( r^2+l_0^2\right)^{5/2}}.\label{density}
\end{equation}

In contrast to General Relativity (GR), which is a local theory, here we aim to incorporate an additional term (i.e., the self-energy associated with the gravitational field) that arises as a non-local effect, leading to a new contribution in the field equations. 
Specifically, we consider a modified, non-local theory of gravity recently proposed in Ref.~\cite{Jusufi:2025qgd}, where it was shown that a non-local gravitational effect leads to the gravitational self energy
 \begin{equation}
 E^{\text{GSE}} = -\frac{1}{2} \int V_G(\textbf{x})\, \rho^{\text{bare}}(\textbf{x})\, d^3 \textbf{x}.
 \end{equation}

This relation plays the role of the Hamiltonian self-interaction $ H^{\text GSE} =  E^{\text{GSE}} $.
If we add this interaction term to the  Schr\"odinger equation, we obtain 
 \begin{equation}
i \hbar \frac{\partial \Psi }{\partial t}=  \left[-\frac{\hbar^2}{2 M} \nabla^2  - \frac{1}{2} \int V_G(\textbf{x})\, \rho^{\text{bare}}(\textbf{x})\, d^3 \textbf{x} \right]\Psi. \label{SNEI}
\end{equation}
Using the fact that the mass density of the source is related to the quantum probability density via 
 \begin{eqnarray}\label{bare1}
     \rho^{\text{bare}}(\textbf{x},t)=M |\Psi(t,\textbf{x})|^2,
 \end{eqnarray}
 we obtain the Schr\"odinger--Newton equation in T-duality with gravitational self-energy corrections 
 \begin{equation}
i \hbar \frac{\partial \Psi }{\partial t}= \left[ -\frac{\hbar^2}{2 M} \nabla^2 -\frac{1}{2}GM^2 \int \, \frac{|\Psi(t,\textbf{x})|^2}{\sqrt{|\textbf{x}-\textbf{y}|^2+l_0^2}}\, d^3 \textbf{x}~ \right] \Psi \,.\label{SNEI2}
\end{equation}

Thus, we have obtained corrections to the Schr\"odinger--Newton equation arising from the non-local gravitational energy. In fact, this modified form of the Schr\"odinger--Newton equation with a zero-point length coincides with the one proposed in \cite{Jusufi:2023tdn}. However, in \cite{Jusufi:2023tdn}, that form was assumed, without providing underlying arguments for such corrections. In the limit $l_0 = 0$, we recover the standard Schr\"odinger–Newton equation studied by Di\'osi and Penrose \cite{Diosi:1986nu,Penrose:1996cv}. However, for $l_0 > 0$, the potential is everywhere finite, removing the $1/r$ singularity. This regularization is not merely a mathematical trick; it encodes the fundamental non-locality of spacetime expected from quantum gravity. We will write this equation in the Newtonian limit. Assuming spherical symmetry to simplify the calculations, we can use Gauss’s law for gravity \cite{Jusufi:2025qgd}
\begin{equation}\label{bare2}
\nabla \cdot \mathbf{g} = -4\pi G \rho^{\text{bare}},
\end{equation}
to obtain
\begin{eqnarray}
 E^{\text{GSE}}  &=& \frac{1}{8\pi G} \int V_G(r)\, (\nabla \cdot \mathbf{g})\, d^3\mathbf{r}. \nonumber \\
 &=& - \frac{1}{8\pi G} \int \mathbf{g} \cdot (\nabla V_G)\,d^3\mathbf{r}
\end{eqnarray}
where we have applied the divergence theorem to arrive at the last result. 
The Newtonian gravitation field is given by
\begin{eqnarray}
\label{new-g}
    \textbf{g}=-\nabla V_G(r)= -\frac{GMr}{(r^2 + l_0^2)^{3/2}}\hat{r}.
\end{eqnarray}
Using \eqref{new-g}, we obtain
\begin{equation}\label{SHN}
i \hbar  \frac{\partial \Psi_N(t,\textbf{r}) }{\partial t}= \left[ -\frac{\hbar^2}{2 M} \nabla^2 + \frac{1}{8\pi G} \int_0^r \mathbf{g}^2(r') \, d^3\mathbf{r}'  \right]\Psi_N(t,\textbf{r}),
\end{equation}
which is the regularized Schr\"odinger--Newton equation in the Newtonian limit. The T-duality--inspired Schr\"odinger--Newton equation given in Eq.~\eqref{SHN}, to the best of our knowledge, is new and has not been reported previously in the literature. We stress that the non-local modification of the gravitational interaction employed here is not introduced ad hoc. Rather, it follows from path-integral duality and T-duality considerations, which imply the existence of a minimal length scale and lead to a UV-regular propagator. The resulting gravitational potential is therefore a natural consequence of quantum gravitational corrections and has been extensively studied in the context of black hole physics and non-singular geometries. In this sense, the regularized potential used in Eq. (2) provides a physically motivated extension of the Newtonian interaction that incorporates short-distance non-local effects.
%%%%%%%%%%%%%%%%%%%%%%%%%%%%%%%%%%%%%%%%%%%%%%%%%%%%%%%%%%%%%%%%%%	
\subsection{Non-linearity of the Schr\"odinger--Newton Equation}\label{subsectionIII}
%%%%%%%%%%%%%%%%%%%%%%%%%%%%%%%%%%%%%%%%%%%%%%%%%%%%%%%%%%%%%%%%%%	

	One of the most striking features of the Schr\"odinger equation in standard quantum mechanics is its linearity.  
	Linearity ensures that if $\Psi_{1}$ and $\Psi_{2}$ are solutions, then any linear combination
	\begin{equation}
    \label{superposition}
		\Psi(t,\mathbf{r}) = c_{1}\Psi_{1}(t,\mathbf{r}) + c_{2}\Psi_{2}(t,\mathbf{r}),
	\end{equation}
	with $c_{1},c_{2}\in\mathbb{C}$, is also a valid solution. This property underlies the superposition principle and is essential for the standard probabilistic interpretation of quantum mechanics.
	
	However, once gravitational self-energy effects are incorporated, as in the modified Schr\"odinger--Newton equation obtained in Eq.~\eqref{SHN}, the situation changes drastically.  
	We now provide an explicit derivation of how the non-linearity arises.
	
%%%%%%%%%%%%%%%%%%%%%%%%%%%%%%%%%%%%%%%%%%%%%%%%%%%%%%%%%%%%%%%%%%	
\subsubsection{Absence of non-linearity in the free case}
%%%%%%%%%%%%%%%%%%%%%%%%%%%%%%%%%%%%%%%%%%%%%%%%%%%%%%%%%%%%%%%%%%

	Neglecting the gravitational self-energy term reduces Eq.~\eqref{SHN} to the ordinary free Schr\"odinger equation
	\begin{equation}
		i\hbar \frac{\partial \Psi_N(t,\mathbf{r})}{\partial t}
		= -\frac{\hbar^2}{2M} \nabla^2 \Psi_N(t,\mathbf{r}),
	\end{equation}
	which is manifestly linear. In this case, the superposition principle \eqref{superposition} holds exactly for $\Psi_N(t,\mathbf{r})$.
%	\begin{equation}
%		\Psi_N(t,\mathbf{r}) = 
%		c_1 \Psi_{1N}(t,\mathbf{r}) + c_2 \Psi_{2N}(t,\mathbf{r}).
%	\end{equation}

%%%%%%%%%%%%%%%%%%%%%%%%%%%%%%%%%%%%%%%%%%%%%%%%%%%%%%%%%%%%%%%%%%	
\subsubsection{Non-linearity from gravitational self-interaction}
%%%%%%%%%%%%%%%%%%%%%%%%%%%%%%%%%%%%%%%%%%%%%%%%%%%%%%%%%%%%%%%%%%

	When gravitational self-energy is included, combining Eqs.~(\ref{bare1}) and (\ref{bare2}), the Newtonian gravitational field is sourced by the quantum probability density,
	\begin{equation}
		\nabla \cdot \mathbf{g} = -4\pi G\, M |\Psi_N(t,\mathbf{x})|^2,
		\label{eq:gDependsOnPsi}
	\end{equation}
	so that the gravitational field $\mathbf{g}$ is no longer an external classical field but a functional of the wavefunction itself.  \\
	 Using Eq.~\eqref{eq:gDependsOnPsi}, we can show that the gravitational potential energy depends on the  wavefunction 
    \begin{equation}
    \label{E-gse}
 E^{\text{GSE}}[\Psi_N]  = \frac{1}{8\pi G} \int_0^r V_G(r')\, (\nabla \cdot \mathbf{g})\, d^3\mathbf{r'}. 
\end{equation}
 Using the divergence theorem, Eq.~\eqref{E-gse} for the gravitational self-energy becomes
	\begin{equation}
		E^{\rm GSE}[\Psi_N]
		= \frac{1}{8\pi G}
		\int_0^{r} \mathbf{g}^2(r')\, d^3\mathbf{r}',
		\label{eq:PhiPsiDefinition}
	\end{equation}
	which is again a nonlinear functional of $|\Psi_N|^2$.
	
	Thus, for two distinct wavefunctions $\Psi_{1N}$ and $\Psi_{2N}$, the functional $\Phi$ does not satisfy a linearity condition:
	\begin{equation}
		E^{\rm GSE}[c_1\Psi_{1N} + c_2\Psi_{2N}]
		\;\neq\;
		c_1\,E^{\rm GSE}[\Psi_{1N}] + c_2\,E^{\rm GSE}[\Psi_{2N}].
	\end{equation}
	This shows that the corresponding Schr\"odinger--Newton evolution equation,
	\begin{equation}
		i\hbar \frac{\partial \Psi_N }{\partial t}
		= \left[
		-\frac{\hbar^2}{2M} \nabla^2
		+ E^{\rm GSE}[\Psi_N]
		\right]\Psi_N,
	\end{equation}
	does not preserve linear superpositions {\it i.e.}, gravity introduces a non-linear term in the quantum dynamics, and this non-linearity breaks the standard quantum superposition principle.

%%%%%%%%%%%%%%%%%%%%%%%%%%%%%%%%%%%%%%%%%%%%%%%%%%%%%%%%%%%%%%%%%%
\subsubsection{Physical implications}
%%%%%%%%%%%%%%%%%%%%%%%%%%%%%%%%%%%%%%%%%%%%%%%%%%%%%%%%%%%%%%%%%%

The breakdown of linearity is not merely a mathematical subtlety, but a profound conceptual shift with potentially far-reaching physical consequences.  
In standard quantum mechanics, the linearity of the Schr\"odinger equation guarantees that the space of solutions forms a Hilbert space: any coherent superposition of solutions is again a solution, and interference phenomena follow directly from this structure. The gravitational self-interaction term in the Schr\"odinger--Newton equation violates this basic property by introducing a wave function-dependent potential. This modification significantly alters not only the mathematical structure of the theory but also the interpretation of quantum states in regimes where gravity cannot be neglected.

In particular, the presence of gravitational self-interaction entails the following:
\begin{itemize}
	\item \textbf{State-dependent evolution and departure from linear unitarity.}  
	The inclusion of the gravitational self-energy term means that the Hamiltonian becomes a functional of the wavefunction. Thus, the time evolution of $\Psi$ is no longer determined by a linear operator acting on the state but instead depends explicitly on $|\Psi|^2$. This leads to a nonlinear evolution equation of the form  
	\[
	i\hbar \frac{\partial \Psi}{\partial t} = \hat{H}[\Psi]\, \Psi,
	\]
	where $\hat{H}[\Psi]$ changes as the state evolves. Such a departure from linearity is incompatible with the superposition principle and, in turn, with the standard interpretation of quantum measurement and decoherence. Moreover, nonlinear quantum dynamics generically do not preserve the norm or the probabilistic structure unless additional constraints are imposed.
	
	\item \textbf{Suppression of macroscopic superpositions and connection to gravitational collapse models.}  
	A central implication of the Schr\"odinger--Newton non-linearity is its natural tendency to suppress spatially extended macroscopic superpositions. This behaviour closely parallels the ideas independently proposed by Di\'osi and Penrose \cite{Diosi:1986nu, Diosi:1989hlx, Penrose:1996cv, Penrose:1998dg}, who argued that gravity may serve as an intrinsic mechanism for objective state reduction. In the Di\'osi--Penrose (DP) framework, the key insight is that a quantum superposition of distinct mass distributions corresponds to a superposition of spacetime geometries; such geometries cannot coexist stably due to the fundamentally nonlinear nature of gravitation. The resulting ``gravitational self-energy'' associated with the difference between the mass configurations sets a characteristic timescale for the decay of the superposition.

	The Schr\"odinger--Newton equation provides a concrete dynamical realization of this concept. Because the gravitational potential is sourced by $|\Psi|^2$, a superposed state does not evolve as a mere linear combination of its components. Instead, each branch of the superposition gravitates, and the nonlinear coupling introduces an effective self-focusing that favours localized states over widely separated ones. For sufficiently massive systems, this effect can mimic the collapse postulated in the DP models, offering a deterministic counterpart to their stochastic collapse dynamics. In this way, the Schr\"odinger--Newton approach not only illustrates how gravity can discourage macroscopic superpositions but also provides an explicit mechanism by which such suppression emerges directly from the gravitational self-interaction of the quantum state.
	
	\item \textbf{Non-linearity even in the presence of T-duality corrections.}  
	The introduction of a zero-point length $l_0$, regularizes the Newtonian potential and removes ultraviolet divergences. However, this modification does not alter the nonlinear structure of the theory. Even with T-duality corrections, the gravitational field remains sourced by $|\Psi|^2$, and the potential retains its dependence on the instantaneous mass distribution encoded in the wavefunction.  
    
    It is worth emphasizing that the nonlinearity encountered here is not an artifact of approximation, but rather a direct consequence of coupling the quantum state to its own gravitational field. In particular, the Hamiltonian becomes a functional of the wavefunction, and therefore the time evolution cannot be described by a linear operator acting on a fixed Hilbert space. This feature distinguishes the present framework from standard quantum mechanics and aligns it with a broader class of nonlinear modifications motivated by gravitational considerations. In the present case, however, the origin of the nonlinearity is fully specified by the non-local gravitational self-energy, rather than being introduced phenomenologically.
    
	Thus, while the zero-point length smooths the potential and changes quantitative features--such as the strength of localization and the onset scale for nonlinearity--it does not restore linearity.

\end{itemize}

%%%%%%%%%%%%%%%%%%%%%%%%%%%%%%%%%%%%%%%%%%%%%%%%%%%%%%%%%%%%%%%%%%	
\subsection{Energy corrections due to gravitational self energy in the Newtonian limit.}
%%%%%%%%%%%%%%%%%%%%%%%%%%%%%%%%%%%%%%%%%%%%%%%%%%%%%%%%%%%%%%%%%%	

The gravitational self-energy term is proportional to the volume integral of the square of $\textbf{g}$
\begin{eqnarray}\label{GSE0}
 E^{\rm GSE}  &=& \frac{1}{8\pi G}\int_0^r \textbf{g}^2(r') d^3\textbf{r}' \nonumber \\
&=&-\frac{5 GM^2 r^3}{16(r^2+l_0^2)^2}-\frac{3 GM^2 l_0^2 r}{16(r^2+l_0^2)^2} \nonumber  \\
&& +\frac{3  GM^2}{16 l_0} \arctan(\frac{r}{l_0}).
\end{eqnarray}

To compute the average energy density for the self-interaction energy, we use Eq.~\eqref{GSE0} and the relation
\begin{eqnarray}
\label{rho-gse}
\rho^{\rm GSE}  = \frac{1}{4\pi r^2}\frac{\partial}{\partial r}( E^{\rm GSE} ) = \frac{M^2r^2}{8\pi(r^2 + l_0^2)^3}.
\end{eqnarray}

If we further expand \eqref{GSE0} in the region $r\gg l_0$, we get
\begin{eqnarray}\label{GSE}
 E^{\rm GSE}  
=\frac{3 \pi GM^2}{32 l_0}-\frac{1}{2 }\frac{GM^2}{ r}+\dots
\end{eqnarray}
The first term gives the gravitational self-energy correction to the mass, while the second term is the Newtonian gravitational potential.
The Schr{\"o}dinger-Newton equation now becomes
\begin{equation}
i \hbar  \frac{\partial \Psi_N(t,\textbf{r}) }{\partial t}= \left[ -\frac{\hbar^2}{2 M} \nabla^2 + \frac{3 \pi G M^2}{32 l_0} +\frac{1}{2}g_N M r
\right]\Psi_N(t,\textbf{r}),\label{sheq}
\end{equation}
where 
\begin{equation}
g_N=-\frac{GM}{r^2},
\end{equation}
is the Newtonian gravitational acceleration. Thus, there is a correction term due to the gravitational self-energy resulting in
\begin{equation}
\hat{H}\Psi_N(\textbf{r})=-\frac{\hbar^2}{2 M} \nabla^2 \Psi_N(\textbf{r}) + \left[\frac{3 \pi G M^2}{32 l_0}-\frac{1}{2 }\frac{GM^2}{ r} \right]\Psi_N(\textbf{r})~.
\end{equation}
The total energy is now
\begin{equation}\label{energy1}
E(r)=\mathcal{E}+\frac{3 \pi G M^2}{32 l_0}-\frac{1}{2 }\frac{GM^2}{ r} ~,
\end{equation}
where we used $-\frac{\hbar^2}{2 M} \nabla^2 \Psi_N(\textbf{r})= \mathcal{E} \Psi_N(\textbf{r})$.
This expression shows that the bare energy of the particle is modified by the regularized gravitational self-energy.

It is important to emphasize the conceptual status of Eq.~\eqref{sheq} in light of the broader discussion on non-linearity. 
At first glance, the Schr\"odinger equation in the Newtonian limit appears to regain linearity, since the additional terms in the Hamiltonian,
\[
\frac{3\pi G M^2}{32 l_0} \qquad \text{and} \qquad \frac{1}{2} g_N M r,
\]
do not depend explicitly on $|\Psi|^2$. This might suggest that the superposition principle is restored. However, this apparent linearity is not fundamental: it arises only because Eq.~\eqref{sheq} is written after imposing a Newtonian approximation in which the gravitational field is treated as an externally specified function of $r$ rather than as a functional of the wavefunction.

In the full formulation of the Schr\"odinger--Newton equation, the gravitational potential is sourced by the probability density $M|\Psi|^2$, making the evolution inherently nonlinear. The Newtonian-limit expression used here corresponds to evaluating the gravitational field for a static, spherically symmetric mass distribution and substituting its approximate asymptotic form back into the Schr\"odinger equation. This procedure removes the explicit $\Psi$-dependence while retaining the dominant energy contribution of the regularized self-gravitational field. As a result, Eq.~\eqref{sheq} captures the leading-order energy corrections due to gravitational self-interaction, but it does not represent the full dynamical structure of the nonlinear theory.

Thus, the linear-looking form of Eq.~\eqref{sheq} is a consequence of the approximation and should not be interpreted as a restoration of fundamental linearity. Instead, it highlights a key lesson:  
the nonlinear features of self-gravitating quantum dynamics may become hidden when one works in limiting regimes, even though they remain essential in the underlying theory. This reinforces the need to analyze the complete Schr\"odinger--Newton equation when discussing superposition, state dependence, and the possible role of gravity in modifying quantum mechanics.

%%%%%%%%%%%%%%%%%%%%%%%%%%%%%%%%%%%%%%%%%%%%%%%%%%%%%%%%%%%%%%%%%%
\section{Wave function collapse}\label{SectionIII}
%%%%%%%%%%%%%%%%%%%%%%%%%%%%%%%%%%%%%%%%%%%%%%%%%%%%%%%%%%%%%%%%%%

As argued in the previous section, gravitational self-energy generally forbids exact linear superpositions. In this sense, standard quantum mechanics without gravitational effects should be viewed only as an approximation theory. In this section, we invert the argument: we assume the validity of the superposition principle and show that such superpositions must become unstable once gravitational effects are taken into account. This reasoning follows Penrose’s argument, which demonstrates a fundamental conflict between the superposition principle and gravity that results in a collapse of the wave
function due to gravitational effects. In this section, we will study the wave function collapse of a superposition state and an entangled state.

%%%%%%%%%%%%%%%%%%%%%%%%%%%%%%%%%%%%%%%%%%%%%%%%%%%%%%%%%%%%%%%%%%
\subsection{Wave function collapse of superposition state}
%%%%%%%%%%%%%%%%%%%%%%%%%%%%%%%%%%%%%%%%%%%%%%%%%%%%%%%%%%%%%%%%%%

 Let us consider an object (for example, a rock) in a superposition. Following \cite{Penrose:2014nha,Howl:2018qdl}, a well-defined quantum state of this object is given by 
\begin{eqnarray}
    \ket{\rm Rock}=c_1 \ket{R_1}+c_2 \ket{R_2},
\end{eqnarray}
where $\ket{R_1}$ and $\ket{R_2}$ are ket vectors of the rock's position in two locations. This leads to a situation where there is a superposition of a gravitational field corresponding to the acceleration $\textbf{g} _1$ and one corresponding to the acceleration $\textbf{g} _2$. This creates a problem when using the Einsteinian perspective; we end up with an ill-defined metric. In Penrose's proposal~\cite{Penrose:2014nha,Howl:2018qdl}, the spacetime background is assumed to be semi-classical. 
In contrast, we suggest that by invoking T-duality, one can obtain a non-local theory characterized by a regular gravitational potential, 
where the metric becomes non-defined at very short distances. 
In other words, a fundamental uncertainty exists in the structure of spacetime at these scales. 
In particular, we can write
 \begin{eqnarray}
 \Delta g =g(r+\Delta r)-g(r)\simeq g'(r) \Delta r,
 \end{eqnarray}
where $\Delta r$ represents the uncertainty in position and $\Delta g$ denotes the uncertainty in acceleration. 
Thus, we obtain 
 \begin{eqnarray}
 \Delta g \simeq \frac{G M (2 r^2-l_0^2)}{(r^2+l_0^2)^{5/2}} \Delta r.
 \end{eqnarray}
We now estimate the following relation
 \begin{eqnarray}
 \frac{\Delta g}{\left| g \right|} = \frac{(2 r^2-l_0^2)}{r(r^2+l_0^2)} \Delta r.
 \end{eqnarray}
The case $\Delta g \ll \left| g \right|$ is not of particular interest to us. Taking $ l_0 \sim l_{Pl}$, we can assume that $r \gg l_0$ and $\Delta r \gg l_0$. The uncertainty is small and of the same order 
  \begin{eqnarray}
  \label{drless}
 \frac{\Delta g}{\left| g \right|} = \frac{(2 r^2-l_0^2)}{r(r^2+l_0^2)}\Delta r\sim \frac{2}{r} \Delta r \ll 1,
 \end{eqnarray}
indicating that $\Delta r \ll r/2$, which applies in the regime of classical gravity. The key step in our derivation is the identification of the regime $\Delta g \sim |g|$. In standard quantum mechanics, the uncertainty $\Delta r$ is independent of $r$. However, in a theory with a minimal length $l_0$, when $r$ approaches $l_0$, the spacetime metric itself becomes ill-defined. We propose that this breakdown occurs when the uncertainty in the gravitational acceleration is comparable to the acceleration itself. In other words, the conflict between the equivalence principle and superposition is shown to originate from the regime $\Delta g \sim |g|$. 

We now consider the case $\Delta g \sim |g|$, 
that is, when quantum effects become significant at distance scales comparable to the quantum uncertainty in position, namely, at very short distances. 
Again, we have $l_0 \sim l_{\mathrm{Pl}}$, but now $r \gtrsim l_0$ and $\Delta r \gtrsim l_0$. 
In this regime, the uncertainty is of the same order as the characteristic length scale of the system, so that quantum gravitational effects cannot be neglected. In contrast to the result in \eqref{drless} leading to 
  \begin{eqnarray}
 \frac{\Delta g}{\left| g \right|}= \frac{(2 r^2-l_0^2)}{r(r^2+l_0^2)} \Delta r \sim \frac{2}{r} \Delta r \sim 1,
 \end{eqnarray}
 with the result that $\Delta r \sim r/2$, which is in contrast to the result  $\Delta r \ll r/2$ coming from \eqref{drless}.
 
Using the condition $\Delta g \sim \left| g \right|$, the Schr\"odinger--Newton equation \eqref{SHN} reads
\begin{equation}
i \hbar  \frac{\partial \Psi_N(t,\textbf{r}) }{\partial t}= \left[ -\frac{\hbar^2}{2 M} \nabla^2 + \frac{1}{8\pi G} \int_0^r (\Delta \mathbf{g})^2 d^3\mathbf{r}'  \right]\Psi_N(t,\textbf{r}).
\end{equation}
 We note here that Penrose defined the gravitational self-energy for the quantum system in a superposition as follows \cite{Penrose:2014nha,Howl:2018qdl}
\begin{eqnarray}
 E^{\rm GSE}_{\rm Penrose}  &=& \frac{1}{8\pi G}\int_0^r (\textbf{g}_2(r')-(\textbf{g}_1(r'))^2 d^3\textbf{r}',
\end{eqnarray}
which is an {\it ad hoc} definition in order to compute the collapse time. In our case, this relation follows naturally from the modified Schr\"odinger--Newton equation along with the condition $\Delta g \sim \left| g \right|$. In other words, the spacetime metric here is not defined in this small length-scale.

The gravitational self-energy term is then proportional to the volume integral of the square of $\Delta\textbf{g}$ given by
\begin{eqnarray}
	E^{\rm GSE}  & \simeq & \frac{1}{8\pi G}\int_0^r (\Delta \mathbf{g}(r'))^2 d^3\textbf{r}'  \nonumber  \\
	&=&  - \frac{\left(389 r^6 + 573 l_0^2 r^4 + 123 l_0^6\right)G M^2 r}{1024 (l_0^2 + r^2)^4} \nonumber \\
	&&  +\frac{123 G M^2 }{1024 l_0}\arctan\left(\frac{r}{l_0}\right).
\end{eqnarray}
%
%\begin{eqnarray}
% E^{\rm GSE}  & \simeq & \frac{1}{8\pi G}\int_0^r (\Delta \mathbf{g}(r'))^2 d^3\textbf{r}'  \nonumber  \\
%&&  \hspace{-1.0cm} = - \frac{389 G M^2 r^7}{1024 (l_0^2 + r^2)^4}-\frac{573 G M^2 l_0^2 r^5}{1024 (l_0^2 + r^2)^4} \nonumber \\
%&& \hspace{-0.5cm} - \frac{123 G M^2 l_0^6 r}{1024 (l_0^2 + r^2)^4} %+\frac{123 G M^2 }{1024 l_0}\arctan\left(\frac{r}{l_0}\right).
%\end{eqnarray}
%
where we used $\Delta r \sim r/2$. If we further expand in the region $r \gtrsim l_0$, we get
\begin{eqnarray}\label{GSE2}
 E^{\rm GSE}  
 \simeq \frac{123 \pi GM^2}{2048 l_0}-\frac{1}{2 }\frac{GM^2}{ r}+\dots
\end{eqnarray}
which is similar in leading order terms with Eq. \eqref{GSE}, note that this discrepancy is due to the fact that they are being solved in a different regime, and as a result, they should not be exactly the same. We emphasize that Eq.~\eqref{GSE2} is an asymptotic expansion valid for $r \gtrsim l_0$. The full expression contains additional terms that are subleading in $l_0/r$. The constant term $123\pi GM^2/(2048 l_0)$ is a result of non-local correction to the self-energy; it does not appear in the standard DP model.
In this region, the Schr\"odinger--Newton equations reads
\begin{equation}
i \hbar  \frac{\partial \Psi_N(t,\textbf{r}) }{\partial t}= \Big[-\frac{\hbar^2}{2 M} \nabla^2+\frac{123 \pi GM^2}{1024 l_0} + M \, g_N \Delta r \Big] \Psi_N(t,\textbf{r}),\label{SHNp}
\end{equation}
where we used the fact that 
\begin{equation}
-\frac{1}{2 }\frac{GM^2}{ r}=-M \left(\frac{GM}{ r^2} \right) \frac{r}{2 }=M \, g_N \Delta r.
\end{equation}
Equation \eqref{SHNp} describes a quantum system when studied from the \textit{Newtonian perspective}, which consists of working in an inertial frame with spacetime coordinates $(\mathbf{r}, t)$ \cite{Penrose:2014nha,Howl:2018qdl}.  However, since $\Delta g_N \sim g_N$,  in our case $\Delta r \sim r/2$, we can write
\begin{equation}
g_N \Delta r \sim\Delta g_N \, (r/2) \sim \Delta \bar{g}_N\, r
\end{equation}
where we defined $ \bar{g}_N=g_N/2$. Due to the constant term 
\begin{equation}
V_0=\frac{123 \pi G M^2}{2048 l_0},
\end{equation}
we obtain a phase factor in the wave function
\begin{equation}
\Psi_N(t,\textbf{r}) =e^{-\frac{i V_0 t}{\hbar}}\tilde{\Psi}_N(t,\textbf{r}).
\end{equation}
The derivation of the phase shift follows Penrose's argument \cite{Penrose:2014nha} but with two important modifications: (i) we use the regularized acceleration $\Delta \bar{g}_N = g_N/2$ obtained from the condition $\Delta g \sim g$, and (ii) we include the constant phase $e^{-i V_0 t/\hbar}$ arising from the non-local self-energy. This constant phase has no observable consequences for probability distributions but becomes relevant when considering interference between different branches of a superposition. Finally, we can write the Schr\"odinger--Newton equation as
\begin{equation}
i \hbar  \frac{\partial \tilde{\Psi}_N(t,\textbf{r})  }{\partial t}= \Big[-\frac{\hbar^2}{2 M} \nabla^2+ M  \Delta \bar{g}_N  r  \Big] \tilde{\Psi}_N(t,\textbf{r}).
\end{equation}

We now follow the argument made by Penrose \cite{Penrose:2014nha} and, more recently, in~\cite{Howl:2018qdl}. 
Penrose proposed a new and elegant way to justify the need for gravity-induced collapse, based on resolving the tension between the superposition principle and the equivalence principle when the gravitational acceleration $\mathbf{g}$ is taken into account. 
 According to the equivalence principle, one may choose to move into a freely falling reference frame, which Penrose calls the \textit{Einsteinian perspective}, where \cite{Penrose:2014nha,Howl:2018qdl}
\begin{equation}
i \hbar  \frac{\partial \Psi_E(T,\textbf{R}) }{\partial T}= -\frac{\hbar^2}{2 M} \nabla^2 \Psi_E(T,\textbf{R}).
\end{equation}

One has to go to local coordinates using $(\textbf{R}, t)$, but since the system is not relativistic, we can set $t=T$ and it is related to $(\boldsymbol{r}, t)$ by 
\begin{equation}
\textbf{R} = \textbf{r} +\frac{1}{2}  \Delta \bar{\textbf{g}}_N t^2~.
\end{equation}
We can set $\textbf{R} =Z \hat{z} $ and  $\textbf{r} =z \hat{z}$ and assume an almost constant $\Delta \bar{\textbf{g}}_N$. In this frame, one solves the free Schr\"odinger equation and then expresses the results in terms of the inertial coordinates $(\boldsymbol{r}, t)$. The solution $\Psi(\textbf{r}, t)$ that relates the Einsteinian perspective and  the Newtonian perspective are \cite{Penrose:2014nha,Howl:2018qdl}
\begin{equation}
\Psi_E(\textbf{R}, t) = e^{i\frac{M}{\hbar} S(\textbf{r},t)} \tilde{\Psi}_N(\textbf{r}, t).
\end{equation}
where 
\begin{eqnarray}
    S(\textbf{r},t)= \frac{1}{6} \Delta \bar{g}_N^2 t^3 + \Delta \bar{\textbf{g}}_N \cdot \textbf{r} \, t .
\end{eqnarray}
The two wave functions are related by a phase factor, hence they should lead to the same physical predictions, {\it i.e.}, $\left| \Psi_E(\textbf{r}, t) \right|^2=\left| \Psi_N(\textbf{r}, t) \right|^2$.  However,  there is a phase difference between the two branches of the superposition given by
\begin{equation}
\Psi_E(\textbf{r}, t) = e^{\frac{i V_0 t}{\hbar}}e^{i\frac{M}{\hbar}\left( \frac{1}{6} \Delta \bar{\textbf{g}}_N^2 t^3 + \Delta \bar{\textbf{g}}_N  \cdot \textbf{r} \, t\right)}\Psi_N(\textbf{r}, t).
\end{equation}

As in Refs.~\cite{Penrose:2014nha,Howl:2018qdl}, we obtain an additional term in the phase factor arising from the regularized gravitational self-energy. 
Moreover, in our framework, the solution naturally emerges from the non-singular Schr\"odinger–Newton equation, incorporating the effects of non-local gravitational self-energy where gravitational acceleration is inherently not well-defined on short scales. 
Similar to Refs.~\cite{Penrose:2014nha,Howl:2018qdl}, we note that the linear $t$-dependent term in the exponent poses no conceptual difficulties, whereas the $t^3$-proportional term is problematic.  The gravity-induced collapse mechanism resolves this by collapsing the wave function once the term, $ \Delta \bar{\mathbf{g}}_N^2 t^3 $, becomes dominant. 

The collapse time  \cite{Penrose:2014nha,Howl:2018qdl} can be estimated by 
\begin{equation}
\label{t-collapse}
\tau_{\rm collapse} \sim \frac{\hbar}{\Delta E^{\rm GSE} }.
\end{equation}

The uncertainty in the gravitational self-energy can be expressed via 
\begin{eqnarray}
\label{delta-GSE}
\Delta  E^{\rm GSE}  
=\left(\frac{\partial    }{\partial r}E^{\rm GSE}\right) \Delta r~.
\end{eqnarray}
 $\Delta  E^{\rm GSE}$, is the uncertainty in the gravitational self-energy. Using Eqs. \eqref{GSE2} and \eqref{delta-GSE}  we obtain 
\begin{eqnarray}
\Delta  E^{\rm GSE}  
= \frac{1}{2 }\frac{GM^2}{ r^2} \Delta r \sim \frac{1}{8 }\frac{GM^2}{ \Delta r }.
\end{eqnarray}
 Using the Planck length and Planck mass relations 
$l_0 \sim l_{Pl}=\sqrt{\frac{\hbar G}{c^3}}$ and $M_{Pl}=\sqrt{\frac{\hbar c}{G}}$,
we get 
\begin{equation}
  \Delta E^{GSE}\sim \frac{1}{8 } \frac{M^2}{M_{Pl}} \frac{l_{Pl}}{\Delta r} c^2.
\end{equation}
    
 Giving the collapse time as
\begin{equation}\label{collpasetime}
    \tau_{\rm collapse} \sim \frac{8  \hbar M_{Pl} \Delta r}{M^2 l_{Pl} c^2 }.
\end{equation}
 The collapse time depends on both the mass of the system and the characteristic length scale of the superposition state. Putting in numerical values for the constants \eqref{collpasetime} becomes
\begin{equation}  \label{timecollapse}
    \tau_{\rm collapse} \sim \frac{1.9 \times 10^{-23} \Delta r}{M^2} [\rm kg^2 s/m].
\end{equation}

  %  \begin{figure}[ht!]
	%	\centering
  %  \includegraphics[scale=0.58]{prob.png}
%\caption{The density plot of the probability of survival in terms of mass $M$ and time $t$ for the case $ \Delta r=10^{-6}$ m.   }
%	\end{figure}

 %   \begin{figure}[ht!]
%		\centering
 %   \includegraphics[scale=0.56]{probd.png}
%\caption{The density plot of the probability of decay in terms of mass $M$ and time $t$ for the case $ \Delta r=10^{-6}$ m.   }
%	\end{figure}

%In Fig.~1, we show the density plot of the collapse time as a function of the mass $M$ and the spatial uncertainty $\Delta r$. 
 As the mass decreases, the collapse time increases, and vice versa. 
It is also important to note that, in our approach, the collapse time is model-independent 
(in the sense that it applies equally to particles, molecules, or other quantum objects), 
and contains only one free parameter, $\Delta r$. 
Crucially, the collapse time is found to be inversely proportional to the squared mass of the quantum system.

In Table I, we have used the equation for collapse time given by Eq. \eqref{timecollapse} and considered three examples of quantum systems in a superposition with $\Delta r=10^{-6}$ m, $\Delta r=10^{-8}$ m and $\Delta r=10^{-10}$ m, and  masses ranging from
$M =10^{-30}$ to $M=10^{-9}$ in units of kg. Given this, in Table I we estimate the value of the collapse time of the wave function for a given mass in superposition  in  seconds.  For particles with small mass (like electrons and protons) the collapse time is very large. This explains why particles with small mass are perfectly good quantum systems in superposition, while for a particle with masses approaching the Planck mass, the system collapses very fast. 

To see this argument more clearly, the state $\ket{\rm Rock}$ spontaneously decays into 
 \begin{eqnarray}
   \ket{\rm Rock}=c_1\ket{R_1}+c_2 \ket{R_2}  \rightsquigarrow  \ket{R_1},
 \end{eqnarray}
 or 
 \begin{eqnarray}
   \ket{\rm Rock}=c_1\ket{R_1}+c_2 \ket{R_2}  \rightsquigarrow  \ket{R_2},
 \end{eqnarray}
respectively.  Following Penrose, we can define the probability of survival  
 \begin{eqnarray}
     P_{\rm survival   }=e^{-t/\tau_{\rm collapse}}=e^{-t \left( \frac{M^2 l_{Pl} c^2 }{8  \hbar M_{Pl} \Delta r}\right)},
 \end{eqnarray}
where we see that with increasing mass, the probability of survival decreases. In a similar way, we can define the probability of decay of the superposition state as
  \begin{eqnarray}
     P_{\rm  decay }=1-e^{-t/\tau_{\rm collapse}}= 1 - e^{-t \left( \frac{M^2 l_{Pl} c^2 }{8  \hbar M_{Pl} \Delta r}\right)}.
 \end{eqnarray}

Here we see that with increasing mass, the probability of decay increases. For very small masses, the survival probability goes to unity, and the decay probability goes to zero, {\it i.e.}, $P_{\rm  survival } \to 1$ and $ P_{\rm  decay } \to 0$, respectively. For large masses, the survival probability goes to zero, and the decay probability goes to unity, {\it i.e.}, $P_{\rm  survival } \to 0$ and $ P_{\rm  decay } \to 1$, respectively. \\
%This fact can be seen from Fig. 2 and Fig. 3. 

\begin{table}[h!]
\centering
\renewcommand{\arraystretch}{1.3}
\begin{tabular}{|c|c|c|c|}
\hline
$M$ [kg] & $\tau_{\rm collapse}$ [s] & $\tau_{\rm collapse}$ [s] & $\tau_{\rm collapse}$ [s] \\
& $\Delta r=10^{-6}$ [m] & $\Delta r=10^{-8}$ [m] & $\Delta r=10^{-10}$ [m] \\
\hline
$10^{-30}$ & $1.9\times10^{31}$ & $1.9\times10^{29}$ & $1.9\times10^{27}$ \\
\hline
$10^{-28}$ & $1.9\times10^{27}$ & $1.9\times10^{25}$ & $1.9\times10^{23}$ \\
\hline
$10^{-26}$ & $1.9\times10^{23}$ & $1.9\times10^{21}$ & $1.9\times10^{19}$ \\
\hline
$10^{-24}$ & $1.9\times10^{19}$ & $1.9\times10^{17}$ & $1.9\times10^{15}$ \\
\hline
$10^{-22}$ & $1.9\times10^{15}$ & $1.9\times10^{13}$ & $1.9\times10^{11}$ \\
\hline
$10^{-20}$ & $1.9\times10^{11}$ & $1.9\times10^{9}$ & $1.9\times10^{7}$ \\
\hline
$10^{-18}$ & $1.9\times10^{7}$ & $1.9\times10^{5}$ & $1.9\times10^{3}$ \\
\hline
$10^{-16}$ & $1.9\times10^{3}$ & $1.9\times10^{1}$ & $1.9\times10^{-1}$ \\
\hline
$10^{-14}$ & $1.9\times10^{-1}$ & $1.9\times10^{-3}$ & $1.9\times10^{-5}$ \\
\hline
$10^{-12}$ & $1.9\times10^{-5}$ & $1.9\times10^{-7}$ & $1.9\times10^{-9}$ \\
\hline
$10^{-10}$ & $1.9\times10^{-9}$ & $1.9\times10^{-11}$ & $1.9\times10^{-13}$ \\
\hline
$10^{-9}$  & $1.9\times10^{-11}$ & $1.9\times10^{-13}$ & $1.9\times10^{-15}$ \\
\hline
\end{tabular}
\caption{We present the numerical values for the collapse time $\tau_{\text{collapse}}$ using our Eq. \eqref{timecollapse} which is a function of mass $M$ and spatial separation $\Delta r$.}
\end{table}
The condition $\Delta g \sim g$ plays a central role in our analysis and deserves further clarification. This regime corresponds to distances at which the uncertainty in the gravitational acceleration becomes comparable to its mean value, signaling a breakdown of the classical notion of spacetime. In this sense, the gravitational field cannot be treated as a well-defined background, and the notion of a single, coherent spacetime geometry ceases to be meaningful. It is precisely in this regime that the tension between the equivalence principle and the superposition principle becomes unavoidable, leading to an instability of quantum superpositions. Thus, wave function collapse in our framework is not introduced as an additional postulate, but rather emerges as a consequence of the loss of spacetime definiteness induced by non-local gravitational effects.

%%%%%%%%%%%%%%%%%%%%%%%%%%%%%%%%%%%%%%%%%%%%%%%%%%%%%%%%%%%%%%%%%%
\subsection{Wave function collapse of entangled state}
%%%%%%%%%%%%%%%%%%%%%%%%%%%%%%%%%%%%%%%%%%%%%%%%%%%%%%%%%%%%%%%%%%

As we pointed out, the wave function collapse time given by Eq.  \eqref{timecollapse} is induced by a non-local gravitational effect applied  to a quantum superposition state. Here, we will argue that the same collapse time relation applies to  entangled quantum state. Recently, in \cite{Jusufi:2025rlr}, it was shown that the ${\rm ER}={\rm EPR}$ conjecture can be manifested in terms of the Einstein–Rosen wormhole geometry connecting two entangled particles with the gravitational self-energy effect.  In particular, let us assume the existence of a pair created in a maximally entangled state given by
\begin{equation}
\ket {\Psi} = \frac{1}{\sqrt{2}}\left(\ket{R_1 L_2}+\ket{L_1 R_2}\right),
\end{equation}
where the states $\ket{R_1 L_2}$ and $\ket{L_1 R_2}$ give the position of the rock (right rock and left rock) at the position $1$ and position $2$, respectively. The geometry of such a quantum state is connected by an ER bridge with a metric given by  \cite{Jusufi:2025rlr}
\begin{equation}\label{metric2}
    ds_{\pm}^2=-f(u_{\pm})dt^2+\frac{du_{\pm}^2}{\left(1-\frac{l_0^2}{u_{\pm}^2}\right) f(u_{\pm})}+(u^2_{\pm}-l_0^2)\,d\Omega^2,
\end{equation}
with 
\begin{equation}
    f(u_{\pm})=1-\frac{2M}{|u_{\pm}|}\left(1-\frac{l_0^2}{u^2_{\pm}}\right)+\frac{(M^2+Q^2)} {u_{\pm}^2} F(u_{\pm})~.
\end{equation}
 $M$ and $Q$ are the mass and charge of the system and \cite{Jusufi:2025rlr}
\begin{equation}
F(u_{\pm})=\frac{5}{8}-\frac{l_0^2}{4 u_{\pm}^2}-\frac{3 u_{\pm}^2}{8  l_0 \sqrt{u_{\pm}^2-l_0^2}}\arctan\left(\frac{\sqrt{u_{\pm}^2-l_0^2}}{l_0}\right),
\end{equation}
where  $u \in (-\infty, -u_{\rm min}] \cup [u_{\rm min}, \infty)$ and for the zero throat wormhole we have $u_{min}=l_0$. In general, $Q \ll M$, hence the mass dominates the geometry. 
We interpret this geometry mathematically as two congruent parts or ``sheets'', joined by a hyperplane at the wormhole throat. Due to quantum gravity effects, a smooth transition is expected to exist between the wormhole interior and the two external geometries. One can see the total mass/energy of the system if, for large distances $|u_{\pm}| \gg l_0$, we  write 
\begin{equation}
    f(u_{\pm})=1- \frac{2 E(u_{\pm})}{u_{\pm}}+\dots, \label{APPMETRIC}
\end{equation}
where the energy is defined as 
\begin{equation}\label{energy2}
    E(u_{\pm})=M+\frac{3 \pi M^2}{32l_0 }-\frac{M^2}{2u_{\pm}}.
\end{equation}
It is interesting to see that the energy \eqref{energy2} of the wormhole system is exactly the same as the energy of the quantum system predicted by the Schrödinger--Newton equation in \eqref{energy1}. The equivalence between the quantum energy \eqref{energy1} and the wormhole energy \eqref{energy2} is nontrivial. In the quantum calculation, $E(r)$ comes from the Schr\"odinger–Newton equation with the regularized potential. In the wormhole calculation, $E(u_{\pm})$ comes from the asymptotic expansion of the metric function $f(u_{\pm})$ derived from the same T-duality-inspired action. Their agreement supports the ER=EPR conjecture within our non-local gravity framework: the entangled quantum state is geometrically represented by an ER bridge, and the collapse of the wave function corresponds to the instability of the wormhole.

The wormhole configurations are quantum–mechanical objects. Thus, systems with higher mass are expected to be unstable, while systems with  small mass can be stable for longer periods of time. 
In order to compute the characteristic timescale,  we use the uncertainty relation
$
\tau_{\rm collapse} \sim \frac{\hbar}{\Delta E},
$
where $\Delta E$ denotes the energy uncertainty. 
 From Eq. \eqref{energy2}, we get the uncertainty in energy as
$
\Delta E \sim \frac{M^2}{2u_{\pm}^2} {\Delta u_{\pm}}.
$
If we assume that the uncertainty in distance is of the order of the distance, {\it i.e.}, $\Delta u_{\pm} \sim u_{\pm}$, and restoring the physical constants $\hbar$, $c$, and $G$, the instability timescale of the wormhole becomes
\begin{equation}
\tau_{\rm collapse} \sim \frac{\hbar\, M_{\rm Pl} \, u_{\pm}}{M^{2} l_{\rm Pl} c^{2}},
\end{equation}
which agrees with the wave function collapse given by Eq.  \eqref{timecollapse}. Thus, the wave function sources the wormhole geometry, and the collapse of the wormhole is equivalently a collapse of the wave function thus giving a geometric representation of the wave function collapse.

%%%%%%%%%%%%%%%%%%%%%%%%%%%%%%%%%%%%%%%%%%%%%%%%%%%%%%%%%%%%%%%%%%
\section{Recent experimental constraints on Diosi-Penrose model}
\label{SectionIV}
%%%%%%%%%%%%%%%%%%%%%%%%%%%%%%%%%%%%%%%%%%%%%%%%%%%%%%%%%%%%%%%%%%

A recent underground experiment by Donadi \emph{et al.}~\cite{Donadi:2020kzc} placed stringent bounds on a specific stochastic implementation of the Diósi--Penrose (DP) model by searching for spontaneous photon emission from germanium nuclei. Their analysis assumes a Markovian Lindblad master equation (equation (3) in \cite{Donadi:2020kzc}) in which gravitationally induced collapse is driven by a universal white-noise field acting on matter. This noise generates momentum diffusion and, for charged particles, a faint but detectable rate of electromagnetic radiation. The absence of such radiation excludes the parameter-free version of the stochastic DP model built on this master equation.

It is crucial to emphasize that the constraints derived in Ref.~\cite{Donadi:2020kzc} apply \emph{only} to collapse models that employ this specific stochastic noise mechanism. The DP master equation (equation (3) in \cite{Donadi:2020kzc}) is of the Lindblad form,
\begin{equation}
\frac{d\rho}{dt} = -\frac{i}{\hbar}[H,\rho] -\frac{4 \pi G}{\hbar} \int d^3 \textbf{x}\int d^3 \textbf{y}\, \frac{[\hat {M} (\textbf{y}),[\hat {M} (\textbf{x}),\rho]]}{|\textbf{x}-\textbf{y}|},
\end{equation}
with $H$ is the Hamiltonian and $\hat {M}$ gives the total mass density. This noise term generates momentum diffusion and, for charged particles, spontaneous photon emission.

The collapse mechanism developed in the present work is fundamentally different. The evolution law derived in Sec.~II is the deterministic Schr\"odinger–Newton equation \eqref{SNEI2}, which contains no stochastic terms, no Lindblad operators, and no external noise field. Collapse in our model arises from the condition $\Delta g \sim g$ and the resulting dominance of the cubic gravitational phase, not from a master equation with a fixed collapse rate. Consequently, our model predicts neither universal diffusion of charged particles nor spontaneous photon emission and is therefore not subject to the bounds of Ref.~\cite{Donadi:2020kzc}.

We note that an experimental test of our model would require a different setup, such as atom interferometry measuring the gravitationally induced phase shift $\Delta \phi \sim (M/\hbar)(\Delta g)^2 t^3$ for masses approaching the Planck scale, or optomechanical systems probing the transition from linear to nonlinear evolution at the collapse time scale $\tau_{\rm collapse} \sim \hbar/(G M^2/\Delta r)$.

Conceptually, the difference between the DP model constrained in Ref.~\cite{Donadi:2020kzc} and the model in the present work can be stated in the following way: The stochastic DP model uses the gravitational self-energy of the \emph{difference} between mass distributions as an input to the collapse rate in a master equation. In contrast, our model derives collapse from the non-local gravitational self-energy computed from the T-duality--regularized potential and from the emergence of metric indefiniteness at short scales. Although both frameworks involve gravitational self-energy, the dynamical realizations are inequivalent: one is stochastic and radiative, the other deterministic and non-radiative. 

More generally, the distinction between stochastic and deterministic implementations of gravitationally induced collapse is crucial. The experimental bounds obtained in Ref.~\cite{Donadi:2020kzc} apply specifically to models in which collapse is driven by a universal noise field, leading to diffusion and radiation effects. 
In contrast, our framework contains no such stochastic ingredient. The dynamics are governed entirely by a nonlinear but deterministic Schrödinger--Newton equation, in which the collapse timescale emerges from the intrinsic gravitational self-energy and spacetime uncertainty, rather than from coupling to an external noise source. This places the present model outside the class of theories constrained by these experiments, while still maintaining a direct connection to the physical mechanism proposed by Penrose. Despite a numerical similarity between the predicted collapse rates, the two approaches lead to distinct phenomenological consequences. The stochastic Diósi--Penrose (DP) model considered in Ref.~\cite{Donadi:2020kzc} introduces a universal white-noise field that induces spontaneous photon emission from charged particles. By contrast, no such effect arises in our model, where collapse is purely deterministic and originates from nonlinear self-interaction. Therefore, the bounds of Ref.~\cite{Donadi:2020kzc} do not apply to our framework, as we discuss in detail in Sec.~\ref{SectionIV}.

%%%%%%%%%%%%%%%%%%%%%%%%%%%%%%%%%%%%%%%%%%%%%%%%%%%%%%%%%%%%%%%%%%
\section{Conclusion}\label{Section:Conclusion}
%%%%%%%%%%%%%%%%%%%%%%%%%%%%%%%%%%%%%%%%%%%%%%%%%%%%%%%%%%%%%%%%%%

In this work, we have shown that non-local gravitational self-energy---arising naturally
from string-theoretic T-duality---can be consistently incorporated into the Schr\"odinger--Newton
framework. The resulting effective gravitational potential encodes non-local energy stored in the
field and renders spacetime intrinsically non-local at short scales. This modification regularizes
the Schr\"odinger--Newton equation, removing the short-distance divergences of the standard
Penrose--Di\'osi construction and recovering it smoothly in the limit $l_{0}\to 0$. Within this
framework we derived regularized corrections to the gravitational self-energy, which generate a
global gravitationally induced phase shift in the quantum state.

A key outcome of our analysis is that the collapse time derived here does not rely on
phenomenological assumptions typically invoked in the Penrose--Di\'osi model. In conventional
treatments one introduces a Gaussian mass distribution and an ad hoc cutoff radius $R_{0}$ to
control ultraviolet divergences. Here the collapse time emerges \emph{model-independently}, provided
the spatial uncertainty satisfies $\Delta r \sim r/2$. Under this condition, wave-function collapse
follows directly from the incompatibility between linear superposition and the equivalence
principle in the presence of gravity. This supports the view that gravitationally induced collapse
is not an auxiliary postulate, but rather an unavoidable consequence of combining quantum
coherence with a non-local spacetime geometry.

Our formulation also differs conceptually from Penrose’s original argument. Whereas Penrose
assumes a semiclassical but well-defined background geometry that becomes inconsistent when
different branches of a superposition correspond to different accelerations, the inclusion of
non-local gravitational effects leads instead to a breakdown of metric definiteness itself.
Fluctuations satisfying $\Delta g \sim g$ mirror the positional uncertainty $\Delta r \sim r/2$,
indicating that spacetime loses classical meaning at the same scale at which superpositions become
unstable. This correspondence provides a unified interpretation of the collapse mechanism.

The same non-local corrections generate distinct gravitational phases when comparing inertial
and freely falling frames. These include both linear and cubic time-dependent terms, alongside a
global contribution from the regularized self-energy. Such phases have no analogue in standard
quantum evolution and provide an observable imprint of the underlying non-local structure of
spacetime. They also determine a spontaneous collapse timescale that scales inversely with the
square of the mass. Microscopic systems therefore exhibit exceedingly long collapse times, while
macroscopic objects collapse so rapidly that superpositions of distinct mass configurations never
form. This provides a dynamical and conceptually grounded explanation for the quantum-to-classical
transition without invoking environmental decoherence.\\

Several open questions arise naturally from this work. An important next step is to formulate a
fully relativistic version of the non-local Schr\"odinger--Newton equation, potentially through
duality-invariant extensions of semiclassical Einstein equations. Another promising direction is the
search for experimental signatures of the gravitationally induced phases identified here, for
example in atom interferometry or opto-mechanical systems approaching the collapse threshold.
Clarifying the relation between this framework and quantum field theory in curved spacetime,
including its extension to multi-particle and entangled states, also remains an important task.
Ultimately, a deeper understanding of how non-local gravitational effects shape quantum dynamics
may illuminate broader questions, such as the emergence of classical spacetime, the black-hole
information problem, and possible holographic structures underlying quantum gravity.

While the present framework provides a consistent and physically motivated mechanism for gravitationally induced wave function collapse, several open questions remain. In particular, the exact mechanism of collapse or the precise microscopic mechanism underlying the transition from deterministic nonlinear evolution to an effectively probabilistic collapse remains to be fully understood. The Schr\"odinger–Newton equation \eqref{SNEI2} is fully deterministic. However, in the regime $\Delta g \sim g$ identified in Sec.~\ref{SectionIII}, the gravitational acceleration becomes ill-defined, and the metric loses its classical meaning. This introduces an intrinsic uncertainty that cannot be resolved within the semiclassical approximation. Thus, while the evolution is deterministic for $r \gg l_0$, at scales $r \gtrsim l_0$ the theory does not prescribe a unique outcome; rather, it predicts a spontaneous collapse with probability $P_{\rm decay} = 1 - e^{-t/\tau_{\rm collapse}}$. Whether a more fundamental theory (e.g., a full quantum gravity or a hidden-variable extension) could provide a deterministic mechanism for the choice of branch remains an open question. Nevertheless, the results presented here demonstrate that key features of wave function collapse can emerge naturally from non-local gravitational effects, without the need for phenomenological modifications or stochastic assumptions.  Our framework shows that regardless of the underlying mechanism, the timescale for collapse is forced by the non-local structure of spacetime and scales as $1/M^2$, explaining why macroscopic superpositions are never observed.

%%%%%%%%%%%%%%%%%%%%%%%%%%%%%%%%%%%%%%%%%%%%%%%%%%%%%%%%%%%%%%%%%%
\begin{acknowledgments}
	FSNL acknowledges funding from the Fundacão para a Ciência e a Tecnologia (FCT) through the research grants UID/04434/2025 and PTDC/FIS-AST/0054/2021. FSNL also acknowledges support from the FCT Scientific Employment Stimulus contract with reference CEECINST/00032/2018. DS acknowledges support from the Frank Sutton Research Fund and a CSU Fresno RSCA Award.  	
\end{acknowledgments}
%%%%%%%%%%%%%%%%%%%%%%%%%%%%%%%%%%%%%%%%%%%%%%%%%%%%%%%%%%%%%%%%%%

\bibliography{PLB_accepted}

\end{document}